\documentclass[]{article}
\usepackage[onecolumn]{emulateapj}

\def\two{\hbox{$_{2}$}}
\def\three{\hbox{$_{3}$}}
\def\four{\hbox{$_{4}$}}
\def\five{\hbox{$_{5}$}}
\def\six{\hbox{$_{6}$}}
\def\seven{\hbox{$_{7}$}}
\def\eight{\hbox{$_{8}$}}
\def\nine{\hbox{$_{9}$}}

\newcommand{\dVx}{d^3{\bf x}\;}
\newcommand{\dVxp}{d^3{\bf x'}\;}

\lefthead{~}
\righthead{Constructing a Mass-Current Radiation-Reaction Force}

\begin{document}

\title{Constructing a Mass-Current Radiation-Reaction Force For
	Numerical Simulations}

\author{~}
\author{Luciano Rezzolla\altaffilmark{1}, Masaru Shibata\altaffilmark{2}}
\affil{Department of Physics, University of Illinois at
        Urbana-Champaign, IL 61801}
\author{~}

\author{Hideki Asada}
\affil{Faculty of Science and Technology, Hirosaki University, 
	Hirosaki 036-8561, Japan} 
\author{~}

\author{Thomas W. Baumgarte and  Stuart L. Shapiro\altaffilmark{1}}
\affil{Departments of Physics and Astronomy, University of 
	Illinois at Urbana-Champaign, IL 61801}
\author{~}

\altaffiltext{1}{Also at the National Center for Supercomputing
	        Applications, University of Illinois at
	        Urbana-Champaign, IL 61801}

\altaffiltext{2}{Permanent address: Department of Earth and Space
		Science,~Graduate School of Science Osaka
		University,~Toyonaka,~Osaka 560-0043,~Japan}

\begin{abstract}
We present a new set of 3.5 Post-Newtonian equations in which
Newtonian hydrodynamics is coupled to the nonconservative effects of
gravitational radiation emission. Our formalism differs in two
significant ways from a similar 3.5 Post-Newtonian approach proposed
by Blanchet (1993, 1997). Firstly we concentrate only on the
radiation-reaction effects produced by a time-varying mass-current
quadrupole $S_{ij}$. Secondly, we adopt a gauge in which the
radiation-reaction force densities depend on the fourth time
derivative of $S_{ij}$, rather than on the fifth, as in Blanchet's
approach. This difference makes our formalism particularly well-suited
to numerical implementation and could prove useful in performing fully
numerical simulations of the recently discovered \hbox{$r$-mode}
instability for rotating neutron stars subject to axial perturbations.
\end{abstract}


\section{Introduction}
\label{intro}

	The recent discovery of an instability in \hbox{$r$-mode}
oscillations in relativistic rotating stars has generated widespread
interest. \hbox{$r$-mode} oscillations in Newtonian rotating stars
have been widely investigated in the past (see, for instance,
Papaloizou and Pringle 1978; Provost et al. 1981; Saio 1982), but
the evidence that they are indeed unstable to the emission of
gravitational radiation is rather recent. The first numerical
calculations carried out by Anderson (1998) and confirmed analytically
by Friedman and Morsink (1998) have spawned a growing literature on
the subject (Andersson, Kokkotas and Schutz 1998; Kojima 1998;
Kokkotas and Stergioulas 1998; Levin 1998; Lindblom and Ipser 1999;
Lindblom, Mendell and Owen 1999; Lindblom, Owen and Morsink 1998;
Lockitch and Friedman 1998; Madsen 1998; Owen et al. 1998). Much of
the interest in \hbox{$r$-mode} oscillations is related to the fact
that their existence is not dependent on a specific rate of rotation;
these modes are, in fact, unstable for arbitrarily slowly-rotating,
perfect fluid stars. This result represents a significant difference
from previously investigated, rotation-induced instabilities, as for
instance the bar-mode instability, which requires a minimum rotation
rate of the star (Chandrasekhar 1970; Friedman and Schutz 1978;
Lindblom and Mendell 1995; Stergioulas and Friedman
1998). Consequently, the \hbox{$r$-mode} instability may have a more
pervasive effect.

	The \hbox{$r$-mode} instability is a purely relativistic
effect, triggered by the emission of gravitational radiation and can
be explained in terms of the basic Chandrasekhar-Friedman-Schutz
instability mechanism (Chandrasekhar 1970; Friedman and Schutz
1978). Since gravitational radiation removes positive angular momentum
from a prograde mode (i.e. a mode that in an inertial frame is seen as
moving in the same positive $\varphi$ direction as the star), it will
also extract positive angular momentum from any perturbation which (as
a result of the star's rotation) is prograde in the inertial, but
retrograde in the corotating frame. Such a mode has, in the corotating
frame, negative angular momentum (the perturbed fluid does not rotate
as fast as it did without the perturbation) and by making its angular
momentum increasingly negative, gravitational radiation drives an
instability (Friedman 1998). A significant difference between the
axial \hbox{$r$-mode} instability and the previously known
gravitational radiation driven bar-mode instability, is that
gravitational radiation couples with \hbox{$r$-mode} oscillations
primarily through time-varying mass-current multipole moments rather
than through the usual time-varying mass multipole moments. We here
present a new set of Post-Newtonian (PN) hydrodynamical equations
which make the calculation of radiation-reaction forces due to
mass-current multipole moments numerically feasible.

	There are a number of reasons why a numerical investigation of
the onset, growth and saturation of the \hbox{$r$-mode} oscillations
is of great interest. All investigations to date have been based on
perturbation analyses usually truncated at the lowest order in the
expansion parameter, namely the star's angular velocity (Andersson
1998; Friedman and Morsink 1998; but see Andersson, Kokkotas and
Schutz 1998; Kokkotas and Stergioulas 1998; Lindblom, Mendell and Owen
1999 for a treatment at higher order). Within this framework, the
perturbation is parameterized in terms of the angular velocity of the
star and its growth is heuristically followed in terms of the energy
and angular momentum losses produced by the gravitational radiation
emission (Owen et al. 1998). The relevant timescales for the growth
and the subsequent viscous decays of the instability are also
estimated through perturbation analysis (Lindblom, Owen and Morsink
1998). Although neglecting the presence of a magnetic field, these
approaches have clarified the basic features of the instability and
provided the first estimates of the importance of the instability in
extracting angular momentum from hot young neutron stars, thus setting
an upper limit on their angular velocity (Lindblom, Owen and Morsink
1998; Andersson, Kokkotas and Schutz 1998). They also provided
qualitative and quantitative information about the expected
gravitational waveforms (Owen et al. 1998). However, they are not
capable of describing the nonlinear development of the instability or
identifying its point of saturation. For these features, numerical
simulations are required.

	To study this mode numerically, it will be useful and adequate
to follow the conservative Eulerian hydrodynamics via Newtonian
equations and all the nonconservative effects due to (mass-current
multipoles) gravitational radiation emission via a Post-Newtonian
radiation-reaction potential at the $3.5$ order [this is what we shall
refer to as $(0+3.5)$ PN]. Such a $(0+3.5)$ PN approach has the
advantage of capturing of all the relevant nonconservative general
relativistic effects without having to resort to a more complicated
relativistic hydrodynamics treatment. Moreover, a $(0+3.5)$ PN
treatment allows us to clearly disentangle the different sources of
gravitational radiation. As a result, in a simulation of the
\hbox{$r$-mode} instability, we can selectively neglect all the
dissipative contributions coming from mass multipole moments and
concentrate solely on the mass-current quadrupole moment, which we
expect to be the dominant mass-current multipole moment. Disentangling
these modes is not possible in a full general relativistic treatment.
Finally, because simulations of the onset and growth of the
\hbox{$r$-mode} instability also require the numerical evaluation of
stable configurations on growth timescales much longer than the
dynamical timescales (i.e. growth timescales $\gg$ rotation period),
we also expect that a three-dimensional, fully relativistic simulation
may be, at present, beyond reach. On the other hand, a
radiation-reaction formalism, though approximate, allows us to use a
scaling in order to artificially accelerate the onset of an
instability without changing the underlying physical evolution.

	The organization of the paper is as follows: in Section
\ref{be} we summarize the basic steps of a PN expansion and the
hydrodynamical equations that emerge from it. Particular attention
will be paid to the radiation-reaction forces and to the losses they
produce in the energy and angular momentum of the system. In the
following Sections \ref{b_g} and \ref{n_g} we discuss two explicit
expressions for the radiation-reaction force due to time-varying mass
current quadrupole moments. The first one was obtained by Blanchet
(1997) (Section \ref{b_g}), while the second is derived in a new gauge
(Section \ref{n_g}) which makes it better suited for numerical
implementation. Section \ref{n_g} also contains detailed verifications
that the new force yields the required rates of energy and angular
momentum loss. Section \ref{heqs} synthesizes the main results derived
in the previous Sections and, for the benefit of the reader interested
in numerical implementation, presents the final form of the
hydrodynamic equations and radiation-reaction terms. Having in mind
the study of the \hbox{$r$-mode} instability, we present in Section
\ref{timescales} a useful rescaling of the radiation-reaction term
which will accelerate the growth-time of the instability to a
timescale set by numerical constraints. Section \ref{conclusion}
summarizes our conclusions and the prospects for numerical
computations exploiting the formalism introduced here. Throughout the
paper we will adopt Cartesian coordinates. $G$ and $c$ denote the
gravitational constant and the speed of light. Greek indices run from
0 to 3, Latin indices from 1 to 3, and we use Einstein summation
convention on matched indices.

\section{PN Expansion: The Basic Equations}
\label{be}

	In this Section we briefly summarize the standard approach to
perform a PN expansion for the equations of relativistic
hydrodynamics. (A recent discussion of the foundations and
applications of the PN approximation has been given by Asada and
Futamase 1997). We adopt the standard \hbox{$3+1$} splitting of
spacetime and write the line element in the form

\begin{equation}
ds^2 = g_{\mu\nu} dx^{\mu}dx^{\nu} =
	 -(\alpha^2-\beta_i\beta^i) c^2 dt^2+2\beta_i c dt dx^i+
	\gamma_{ij}dx^i dx^j \ ,
\end{equation}
where $\alpha$, $\beta^i$, and $\gamma_{ij}$ are the lapse function,
the shift vector, and spatial 3-metric, respectively, while $\beta_i =
\gamma_{ij} \beta^j$.

	We also consider a perfect fluid, whose energy-momentum tensor
is

\begin{equation}
T^{\mu\nu}=(\rho c^2 +\rho \varepsilon+ P)u^{\mu}u^{\nu}+P g^{\mu\nu} \ ,
\end{equation}
where $\rho$ is the rest-mass density, $\varepsilon$ is specific
internal energy, $P$ the pressure, and $u^{\mu}$ the fluid
four-velocity. It is also convenient to introduce a coordinate velocity
$v^i$, defined as

\begin{equation}
\label{veq}
{v^i \over c}\equiv \frac{u^i}{u^0}=
	-\beta^i+{\gamma^{ij} u_j \over u^0} \ . 
\end{equation}

	Imposing the conservation of rest-mass, energy and momentum
yields the standard relativistic hydrodynamic equations,

\vbox{
\begin{eqnarray}
\label{hydr1} 
&&{\partial \rho_* \over \partial t}+
	{\partial (\rho_* v^i) \over \partial x^i}=0\ ,
\\
\label{hydr2}
&& {\partial (\rho_* \varepsilon ) \over \partial t}
	+{\partial (\rho_* \varepsilon v^i) \over \partial x^i}=-P\biggl[
	{\partial (\alpha u^0 \gamma^{1/2}) \over \partial t}
	+{\partial (\alpha u^0 \gamma^{1/2} v^i) 
	\over \partial x^i}\biggr]\ ,
\\
\label{hydr3}
	&&{\partial (\rho_* h u_k) \over \partial t}+
	{\partial (\rho_* h u_k v^i) \over \partial x^i}
	=-{\alpha \gamma^{1/2} \over c}{\partial P \over \partial x^k}
	+\rho_* h c\biggl[-\alpha u^0 {\partial \alpha \over \partial x^k}
	+u_j {\partial \beta^j \over \partial x^k}
	-{u_i u_j \over 2 u^0} 
	{\partial \gamma^{ij} \over \partial x^k}\biggr]\ ,
\end{eqnarray}}
where

\begin{mathletters}
\begin{equation}
u^0 = \frac{\left(1 + \gamma^{ij}u_i u_j\right)^{1/2}}{\alpha}\ , 
\end{equation}
\begin{equation}
\rho_* \equiv \rho \alpha u^0 \gamma^{1/2} \ , 
	\hskip 2.0 truecm
	h=1+{1 \over c^2}\biggl( \varepsilon+\frac{P}{\rho} \biggr)\ ,
\end{equation}
\end{mathletters}

\noindent and $\gamma = {\rm det}(\gamma_{ij})$. Equations
(\ref{hydr1})--(\ref{hydr3}) will also be referred to as the
continuity, energy and Euler equations, respectively.

	We next proceed to the PN approximation and perform a series
expansion of the metric in the inverse powers of $c$ up to the 3.5 PN
order (Chandrasekhar 1965; Chandrasekhar and Esposito 1969; Asada,
Shibata and Futamase 1996):

\begin{mathletters}
\label{pnexp}
\begin{eqnarray}
\alpha&=&1+{1 \over c^2}\phi+{1 \over c^4}\four\alpha
        +{1 \over c^5}\five\alpha
        +{1 \over c^6}\six\alpha+{1 \over c^7}\seven\alpha
        +{1 \over c^8}\eight\alpha
        +{1 \over c^9}\nine\alpha + O(c^{-10}) \ , 
\\
\beta^i&=&{1 \over c^3}\three\beta^i+{1 \over c^5}\five\beta^i
        +{1 \over c^6}\six\beta^i+{1 \over c^7}\seven\beta^i
        +{1 \over c^8}\eight\beta^i + O(c^{-9}) \ , 
\\
\gamma_{ij}&=&\delta_{ij}\biggl(1-{2 \over c^2}\phi\biggr)+
        {1 \over c^4}\four h_{ij}+{1 \over c^5}\five h_{ij}
       +{1 \over c^6}\six h_{ij}+
        {1 \over c^7}\seven h_{ij} + O(c^{-8}) \ , 
\end{eqnarray}
\end{mathletters}

\noindent where the left subscript $n$ indicates the coefficient of
the $O(c^{-n})$ term in the series expansion, and $\phi$ is the
Newtonian potential. We implicitly adopt the usual PN gauge in which
$\two \alpha=\phi$ and $\two h_{ij} = -2\phi\delta_{ij}$;
(Chandrasekhar 1965; Chandrasekhar and Esposito 1969) and
$\five\alpha$ is a function of time only. Similarly, the four-velocity
is also expanded in terms of $v^i$, and full expressions for this are
given by Chandrasekhar (1965), Chandrasekhar and Esposito (1969),
Asada, Shibata, and Futamase (1996).

	Using (\ref{pnexp}), the 3.5 PN expression of the Euler
equation (\ref{hydr3}) can then be written as

\begin{eqnarray}
\label{ee_35}
{\partial v_i\over \partial t}+ v_j {\partial v_i \over \partial x_j} =
	-\frac{1}{\rho}{\partial P \over \partial x_i}-
	{\partial \phi \over \partial x_i}+
	\frac{1}{\rho}\left(
	{1 \over c^2}F_i^{\rm 1PN}+{1 \over c^4}F_i^{\rm 2PN}
       +{1 \over c^5}F_i^{\rm 2.5PN}+{1 \over c^6}F_i^{\rm 3PN}+
	{1 \over c^7}F_i^{\rm 3.5PN}\right) + O(c^{-8}) \ .
\end{eqnarray}
Hereafter we neglect the higher order difference between contravariant
and covariant components and use the latter only.

	It is convenient to distinguish, in the right hand side of
(\ref{ee_35}), the terms related to dissipative radiation-reaction
effects from those arising from conservative hydrodynamical
stresses. In particular, we define the radiation-reaction force
density\footnote{Hereafter we refer to the radiation-reaction force
densities simply as radiation-reaction forces.}

\begin{equation}
{\mathbf F}^{\rm rr} \equiv {\mathbf F}^{\rm 2.5PN} + {\mathbf F}^{\rm 3.5PN} 
	= {\mathbf F}^{\rm rrm} + {\mathbf F}^{\rm rrc} \ ,
\end{equation}
where ${\mathbf F}^{\rm rrm}$ and ${\mathbf F}^{\rm rrc}$ refer to the
radiation-reaction force due to time-varying mass multipole moments
and mass-current multipole moments, respectively. The general
(slice-independent) expressions for $F_i^{\rm 2.5PN}$ and $F_i^{\rm
3.5PN}$ are given by (Asada and Futamase 1997; Blanchet 1997)

\begin{eqnarray}
\label{f25}
\rho^{-1}F^{\rm 2.5PN}_i &=& - \partial_i(\seven\alpha)
	- \partial_t(\six \beta_i)
	+ v_j \partial_i (\six\beta_j) 
	- v_j \partial_j (\six\beta_i)
\nonumber\\
	&& - \partial_t (\five h_{ij} v_j)
	- \five h_{ij} v_k \partial_k v_j 
\\ \nonumber\\
\label{f35}
\rho^{-1} F^{\rm 3.5PN}_i &=&-\partial_i (\nine\alpha)
	-\partial_t (\eight\beta_i) - \partial_j(\eight\beta_{i})v_j +
	\partial_i (\eight\beta_{j})v_j 
\nonumber\\
	&&-\partial_t (\seven h_{ij} v_j)- v_k\partial_k (\seven h_{ij}v_j)
	+\frac{1}{2}  v_j v_k \partial_i (\seven h_{jk})
	+ \delta F^{\rm 3.5PN}_i( \seven\alpha,~ \six\beta_i,~ \five h_{ij})
	\ ,
\end{eqnarray}
where $\partial_i = \partial/\partial x_i$, $\partial_t =
\partial/\partial t$, and in deriving (\ref{f25}) we have exploited
that $\partial_i[\five \alpha(t)] = 0$. Note that while ${\mathbf
F}^{\rm 2.5PN}$ is dependent {\it only} on the time-varying {\sl mass
quadrupole} moments , ${\mathbf F}^{\rm 3.5PN}$ is, in general,
dependent on time-varying {\sl mass quadrupole} moments, {\sl mass
octupole} moments, as well as on time-varying {\sl mass-current
quadrupole} moments. In particular, in equation (\ref{f35}), we have
symbolically indicated by $\delta {\mathbf F}^{\rm 3.5PN}(
\seven\alpha,~ \six\beta_i,~ \five h_{ij} v_j)$ all of the
contributions coming from mass quadrupole moments.

	The work done and the torque produced by the
radiation-reaction forces ${\mathbf F}^{\rm rr}$ must balance the
energy and angular momentum carried off to infinity by the
gravitational waves. The emission rate of gravitational waves is known
from the multipolar decomposition of the radiation field (Thorne
1980). In the absence of any dissipative mechanism other that
gravitational wave emission, the energy and angular momentum loss
rates can be readily calculated as

\begin{eqnarray}
\label{dE} 
\frac{d E}{dt}&=&\int \dVx  v_i F_i^{\rm rr} \ , 
\nonumber \\ \\
\label{dL}
\frac{d S_i}{dt}&=&\int \dVx \epsilon_{ijk}x_j F_k^{\rm rr} \ ,  
\end{eqnarray}
where $d/dt = \partial_t + v_i \partial_i$ and $\epsilon_{ijk}$ is the
Levi-Civita symbol. The total energy and angular momentum have the
usual Newtonian definitions,

\begin{eqnarray}
E &\equiv&
	\int \dVx  \rho \Bigl( \frac{1}{2} v^2+\frac{1}{2} \phi+
	\varepsilon \Bigr) \ , 
\nonumber \\ \\
S_i &\equiv& 
	\int \dVx \epsilon_{ijk}x_j \rho v_k \ , 
\end{eqnarray}
with $v^2 \equiv v_kv_k$. Equations (\ref{dE}) and (\ref{dL}) will be
used repeatedly in this paper to verify the correctness of the derived
expressions for the radiation-reaction forces.

	So far, our discussion of radiation-reaction forces has been
general and we have not restricted ourselves to a specific physical
configuration and considered a generic gauge in which the expansion
(\ref{pnexp}) is valid. Hereafter, however, we will concentrate on a
PN formulation of the radiation-reaction forces which may prove useful
in a numerical study of the \hbox{$r$-mode} instability. Because the
\hbox{$r$-mode} instability is predominately excited by mass-current
quadrupole moments (cf. Section \ref{intro}), we will neglect any
contribution to the radiation-reaction forces coming from mass
multipole moments and consider ${\mathbf F}^{\rm rrm} = 0 = \delta
{\mathbf F}^{\rm 3.5PN}$. Even with this restriction, the numerical
computation of ${\mathbf F}^{\rm rrc}$ in presently adopted gauges
(Burke 1971, Blanchet 1997) is nontrivial. In the following two
Sections we discuss these complications and offer a way to simplify
them.

\section{Mass-Current Quadrupole Moment Radiation-Reaction: Blanchet's Gauge}
\label{b_g}

	The first expression for the radiation-reaction force due to
time-varying mass-current quadrupole moments was derived by Burke
using a matched asympotics expansion (Burke 1969, 1970, 1971) and
expressed in terms of vector spherical harmonics. His expression,
however, does not yield the required energy and angular momentum
losses (see Walker and Will 1980 for an explanation of the error in
the formulation). More recently, a new complete treatment of the
radiation-reaction and balance equations at 3.5 PN order has been
provided by Blanchet (1997) as an extension of earlier work on
gravitational radiation-reaction forces (Blanchet 1993). Here we
briefly review the key steps necessary for our modified treatment
presented in the next Section. Firstly, the ``canonical form'' of the
linearized metric ${\bar h}^{\mu \nu}_{(1)}$ is constructed in the
harmonic gauge (Thorne 1980). This linearized metric expresses the
linear deviation from the Minkowski metric $\eta^{\mu \nu}$ in a
series expansion in $G$ [i.e. ${\bar h}^{\mu \nu} \equiv \sqrt {-g}
g^{\mu \nu} - \eta^{\mu \nu} = {\bar h}^{\mu\nu}_{(1)} + {\bar
h}^{\mu\nu}_{(2)} + O(G^3)$]. Then, ${\bar h}^{\mu \nu}_{(1)}$ can be
rewritten in terms of two (and only two) sets of time-varying
multipole moments, referred to as the ``mass-type'' and
``current-type'' moments (Thorne 1980; see Appendix B for
details). The contributions to the linearized ${\bar h}^{\mu
\nu}_{(1)}$ coming from radiation-reaction effects are then derived by
taking the half-sum and the half-difference of the retarded and
advanced expressions of the multipole moments, and by studying the
nonlinear corrections by means of a Post-Minkowskian method (Blanchet
1993, 1997). In doing this, an infinitesimal gauge transformation to
the ``generalized Burke-Thorne gauge'' (hereafter, we will refer to it
simply as {\it Blanchet's gauge}) is performed in order to obtain
$h^{\mu \nu}$, a simplified form of the metric (Blanchet 1993, 1997).

	Restricting our attention {\em only} to the radiation-reaction
force produced by a time-varying mass-current quadrupole moment, it is
clear from (\ref{f35}) that we need explicit expressions for the
metric coefficients $\nine\alpha$, $\eight\beta_k$ and $\seven
h_{ij}$. While the last two are known already from the linear term of
${\bar h}^{\mu\nu}$, the first one needs to be obtained through an
iteration involving also the nonlinear terms. As a result of this
iteration, the relevant parts of expanded metric functions are
[cf. equations (3.6) of Blanchet 1997]

\begin{mathletters}
\label{blanchet_mc}
\begin{eqnarray}
\nine\alpha &=& 0 \ ,
\\
\label{beta8}
\eight\beta_i &=& \frac{16\,G}{45}\epsilon_{ijk}x_j x_l S_{kl}^{(5)} \ ,
\\
\seven h_{ij} &=& 0 \ ,
\end{eqnarray}
\end{mathletters}

\noindent where $S_{ij}$ is the Newtonian mass-current quadrupole
moment defined as

\begin{equation}
\label{Sij}
S_{ij}(t) \equiv \int \dVx \epsilon_{kl(i}x_{j)}x_k \rho v_l \ .
\end{equation}
It is useful to remark that $S_{ij}$ is trace-free, i.e. $S_{ij}
\delta^{ij} = 0 $. The right superscript $(n)$ indicates the $n$-th
total time derivative:

\begin{equation}
A^{(n)}(t) \equiv \left( \frac{d}{dt} \right)^n A(t) \ ,
\end{equation}
and $A_{(ij)} = \frac{1}{2}(A_{ij}+A_{ji})$.

	Using (\ref{f35}), the contribution to the 3.5 PN
radiation-reaction force due to a time-varying mass-current quadrupole
moment in Blanchet's gauge is given by

\begin{equation}
\label{blanc_f35}
\rho^{-1} F_i^{\rm rrc}=\frac{16}{45} G \left(
	2 v_j\epsilon_{jil}x_m S_{lm}^{(5)}
	+v_j\epsilon_{jkl}x_k S_{li}^{(5)} 
	-v_j \epsilon_{ikl}x_k S_{lj}^{(5)}
	-\epsilon_{ikl}x_k x_m S_{lm}^{(6)} 
	\right) \ . 
\end{equation}
The validity of expression (\ref{blanc_f35}) can be verified by
computing the energy and angular momentum dissipation rates. In
particular, inserting (\ref{blanc_f35}) into (\ref{dE}), we
immediately obtain

\vbox{
\begin{eqnarray}
\frac{d E}{dt}&=&\frac{16}{45}{G \over c^7}S_{ij}S_{ij}^{(6)} 
\nonumber \\
	&=&-{16 \over 45}{G \over c^7}S_{ij}^{(3)}S_{ij}^{(3)}
	+{16 \over 45}{G \over c^7}{d \over dt}\biggl(S_{ij}S_{ij}^{(5)}
	-S_{ij}^{(1)}S_{ij}^{(4)}+S_{ij}^{(2)}S_{ij}^{(3)}\biggr). 
\end{eqnarray}}
Assuming nearly-periodic motion of the matter field, and averaging
over several periods, we can discard the total time derivative term
and obtain the standard formula of the energy loss due to mass-current
quadrupole radiation (Thorne 1980):

\begin{equation}
\label{dedteq}
\biggl\langle\frac{d E}{dt}\biggr\rangle=
	-\frac{16}{45}{G \over c^7}\langle S_{ij}^{(3)}S_{ij}^{(3)} \rangle \ ,
\end{equation}
where, as usual, 

\begin{equation}
\langle A \rangle \equiv \frac{1}{T} \int^T_0 A(t) dt \ .
\end{equation}
	
	Similarly, by using expression (\ref{blanc_f35}) in equation
(\ref{dL}), we obtain

\vbox{
\begin{eqnarray} 
\label{dS_idt}
\frac{d S_i}{dt}&=&-\frac{16}{45}{G \over c^7} \int \dVx \rho 
	\Bigl[(x_i x_j x_k S_{jk}^{(6)} - |{\bf x} |^2 x_j S_{ij}^{(6)}) +
	2(x_j x_k v_i S_{jk}^{(5)} - x_j x_k v_k S_{ij}^{(5)})
\nonumber\\ 
	&& \hskip 5.0truecm +(x_i
	x_j v_k S_{jk}^{(5)} - |{\bf x} |^2 v_j S_{ij}^{(5)}) -
	\epsilon_{lmn}\epsilon_{ijk} x_j x_m v_l S_{nk}^{(5)} \Bigr] 
	\nonumber \\
	&=&-{16 \over 45}{G \over c^7}\biggl\{\int \dVx \rho 
	\biggl[(x_j x_k v_i - x_i x_j v_k )S^{(5)}_{jk} -
	\epsilon_{ijk} \epsilon_{lmn} x_m x_j v_l
	S_{nk}^{(5)} \biggr] 
\nonumber \\ 
	&&\hskip 5.0truecm +{d \over dt}\Bigl[ \int
	\dVx \rho (x_i x_j x_k S^{(5)}_{jk}-|{\bf x} |^2 x_j S^{(5)}_{ij}
	\Bigr]\biggr\} 
\nonumber \\ 
	&=&-{32 \over 45}{G \over c^7}\epsilon_{ijk}S_{jl}S_{kl}^{(5)} -
	{16 \over 45}{G \over c^7}{d \over dt} \int
	\dVx \rho (x_i x_j x_k S^{(5)}_{jk}-|{\bf x} |^2 x_j S^{(5)}_{ij}), 
\end{eqnarray}}
\noindent
where $|{\bf x} |^2 \equiv x_i x_i$. After averaging (\ref{dS_idt})
over several periods (and taking two integrations by parts), the
formula for the angular momentum loss is

\begin{equation}
\label{djdteq}
\biggl\langle \frac{d S_i}{dt}\biggr\rangle 
	=-\frac{32}{45}{G \over c^7}\epsilon_{ijk} 
	\langle S_{jl}^{(2)} S_{kl}^{(3)} \rangle \ , 
\end{equation}
which is again in agreement with the required expression (Thorne 1980).

	Although Blanchet's formalism is clear and complete [indeed
Blanchet (1993) and (1997), also discusses 3.5 PN radiation-reaction
potentials due to mass quadrupole and octupole moments], it is not
particularly simple for numerical implementation. The reason for this
is evident from expression (\ref{blanc_f35}), in which the
radiation-reaction force depends on a high time derivative of the
mass-current quadrupole moment $S_{ij}$. It is often possible, and
highly convenient in a numerical calculation, to replace some of the
time derivatives of the mass and mass-current multipole moments by
quadratures. This involves combining the continuity and Euler
equations and introducing some supplementary variables (most notably
the partial time derivatives of the Newtonian gravitational potential)
which satisfy elliptic equations (Nakamura and Oohara 1989). Here,
however, obtaining these quadratures is realistic for $S_{ij}^{(3)}$
at most (see Appendix A for a discussion), leaving the higher order
time derivatives to be obtained via finite differencing of
$S_{ij}^{(3)}$. The latter operation can be extremely inaccurate, even
for a numerical scheme which is second order accurate in time and
space, and might introduce numerical instabilities.

	To overcome this difficulty, we employ a different gauge
condition and derive an alternative form of the metric in the next
Section. This alternative leads to radiation-reaction forces which are
dependent only on the fourth time derivative of $S_{ij}$, a
considerable improvement computationally.

\section{mass-current quadrupole radiation-reaction: a new gauge} 
\label{n_g}

	In this Section we adopt a gauge choice different from
Blanchet's to obtain a more desirable form for the radiation-reaction
force. As in Section (\ref{b_g}), we start with the linear metric in
the canonical form ${\bar h}^{\mu \nu}_{(1)}$ (Thorne 1980; Blanchet
1993) and define ${\tilde h}^{\mu \nu}_{(1)} \equiv {\bar h}^{\mu
\nu}_{(1)} - \frac{1}{2}\eta^{\mu \nu} {\bar h}_{(1)}$, where ${\bar
h}_{(1)} \equiv ({\bar h}_{(1)})^{\mu}_{\ \mu}$. In this gauge, the metric
coefficients $\eight \tilde \beta_k$ and $\seven \tilde h_{ij}$ are
(see Appendix B for details)

\begin{mathletters}
\begin{eqnarray}
\label{harmo8}
\eight \tilde \beta_i&=& {\tilde h}^{0i}_{(1)} = 
	-{4 \,G \over 45}\epsilon_{ijk}x_j x_l S^{(5)}_{kl},
\\
\label{harmo7}
\seven \tilde h_{ij}&=& - {\tilde h}^{ij}_{(1)} = 
	-{8 \,G \over 9} x_k \epsilon_{kl(i} S_{j)l}^{(4)} \ .
\end{eqnarray}
\end{mathletters}

	To eliminate the dependence on the fifth time derivative of
$S_{ij}$, we perform an infinitesimal gauge transformation

\begin{equation}
\label{n_g_metric}
h_{\mu\nu}={\tilde h}_{\mu\nu}+\partial_{\nu}\xi_{\mu}+
	\partial_{\mu}\xi_{\nu} \ ,
\end{equation}
where ${\tilde h}^{\mu \nu} \equiv {\bar h}^{\mu \nu} -
\frac{1}{2}\eta^{\mu \nu} {\bar h}$, with ${\bar h} \equiv {\bar
h}^{\mu}_{\ \mu}$. We use the freedom in the gauge transformation to
set $\eight\beta_k$ to zero by choosing

\begin{mathletters}
\label{xiharmo}
\begin{eqnarray}
\xi_0 &=& 0 \ , 
\\
\xi_i &=& \frac{4\,G}{45}\epsilon_{ijk}x_j x_l S_{kl}^{(4)} \ ,
\end{eqnarray}
\end{mathletters}

\noindent which then yields

\begin{mathletters}
\label{our_mc}
\begin{eqnarray}
\label{beta8eq0}
\eight\beta_i&=&0 \ , 
\\
\label{hij7}
\seven h_{ij}&=&-\frac{32\,G}{45} x_k \epsilon_{kl(i} S_{j)l}^{(4)} \ . 
\end{eqnarray}
\end{mathletters}

	We still have not determined $\nine \alpha$, but this can be
done by choosing a time-slice condition. Note, however, that selecting
a specific form for the shift and the spatial 3-metric through
equations (\ref{our_mc}) restricts the set of possible choices for
$\nine \alpha$ (see Appendix C for a discussion). In particular, we
impose the maximal slicing condition, for which the trace of the
extrinsic curvature tensor is set to zero (Arnowitt, Deser and Misner
1962; Smarr and York 1978; Sch\"afer 1983; Blanchet, Damour and
Sch\"afer 1990) and which is compatible with conditions
(\ref{our_mc}). This results in a linear elliptic equation for the
lapse function, whose 3.5 PN approximation is (see Appendix C for
details)

\begin{equation}
\label{eqalp}
\Delta(\nine\alpha)=\partial_i 
	\left(\seven h_{ij} \partial_j \phi \right)\ , 
\end{equation}
where $\Delta$ denotes the flat spatial Laplacian. Introducing a
scalar ``superpotential'' $\chi$ satisfying (Chandrasekhar 1969)

\begin{equation}
\label{chi_constraint}
\Delta \chi = 2 \phi \ ,
\end{equation}
and using the fact that $\partial_i (\seven h_{ij})=0=\Delta (\seven
h_{ij})$, we then obtain

\begin{equation}
\label{alpha9}
\nine \alpha = 
	{1 \over 2} \left(\seven h_{ij} \partial_{ij} \chi \right) \ . 
\end{equation}

The expression for $\nine \alpha$ can be further simplified by solving
(\ref{chi_constraint}) for $\chi$

\begin{equation}
\label{chi_def}
\chi({\bf x}) \equiv -G\int \dVxp \rho({\bf x'}) |{\bf x} -{\bf x'}| \ . 
\end{equation}
When (\ref{chi_def}) is inserted in equation (\ref{alpha9}), it leads
to

\begin{eqnarray}
\label{nalpha9}
\nine \alpha
	&=&-{1 \over 2}G(\seven h_{ij}) \biggl[
	x_i {\partial \over \partial x_j} \int \dVxp 
	{\rho({\bf x'}) \over |{\bf x} -{\bf x'}|}
	+\delta_{ij} \int \dVxp {\rho({\bf x'}) \over |{\bf x} -{\bf x'}|}
	-{\partial \over \partial x_j} \int \dVxp
	{\rho({\bf x'}) {x'\!}_i \over |{\bf x} -{\bf x'}|}\biggr]
\nonumber \\
	&=&{1 \over 2} (\seven h_{ij})
	[x_i \partial_j \phi + \partial_j P_i] \ ,
\end{eqnarray}
where $\seven h_{ij} \delta_{ij}=0$. The vector potential ${\mathbf
P}$ in (\ref{alpha9}) is defined by

\begin{equation}
{\mathbf P}({\bf x}) \equiv G \int \dVxp
	\rho({\bf x'}) \frac{\bf x'}{|{\bf x} -{\bf x'}|} \ ,
\end{equation}
and can be most easily calculated by solving the linear elliptic
equations

\begin{equation}
\label{pi_eq}
\Delta P_i = -4\pi G \rho x_i\ . 
\end{equation}
While solving equations (\ref{pi_eq}) represents an additional
computation, nonexistent in Blanchet's formulation, this integration
is generally not too taxing in a numerical hydrodynamical simulation
which already must solve Poisson's equation for the Newtonian
gravitational potential (Nakamura and Oohara 1989; Oohara and Nakamura
1990, 1991; Shibata, Nakamura and Oohara 1992; Ruffert, Janka and
Sch\"afer 1996).

	Finally, using expressions (\ref{nalpha9}) and (\ref{hij7})
for $\nine \alpha$ and $\seven h_{ij}$ in (\ref{f35}), we obtain the
new gauge expression for the radiation-reaction force due to a
time-varying mass-current quadrupole moment

\begin{equation}
\label{n_g_f35}
\rho^{-1}F_i^{\rm rrc} = -\partial_i (\nine \alpha) -
	\partial_t (\seven h_{ij}v_j) - 
	v_k \partial_k (\seven h_{ij}v_j) +
	\frac{1}{2} v_j v_k \partial_i (\seven h_{jk}) \ .
\end{equation}

	In subsections \ref{e_l} and \ref{am_l}, we verify
(\ref{n_g_f35}) by computing the energy and angular momentum loss
rates. The reader wishing to omit this discussion may proceed directly
to Section \ref{heqs}.

\subsection{Rate of Energy Loss}
\label{e_l}

	We calculate the rate at which the total energy of the system
is lost to radiation, by substituting expression (\ref{n_g_f35}) into
equation (\ref{dE}). The relevant integrals that emerge are

\begin{eqnarray}
\label{ie1}
\int \dVx  \rho v_i \partial_i (\nine\alpha)
	 &=& -\int \dVx  \partial_i (\rho v_i) \nine\alpha 
	 = \int \dVx  \partial_t (\rho) \nine\alpha 
\nonumber \\
	 &=& -\frac{16}{45}G\epsilon_{ijk} S_{kl}^{(4)} 
	\frac{d}{dt} \int\!\!\int \dVx \dVxp \rho({\bf x} )
	\rho({\bf x'}) 
	\frac{x_{i} x_l {x'\!}_j}{|{\bf x} -{\bf x'}|^3} \ , 
\\ \nonumber \\ \nonumber \\ 
\int \dVx  \rho v_i v_j \partial_t (\seven h_{ij}) &=& \frac{32}{45} 
	G \biggl(S_{ij}^{(1)}S_{ij}^{(5)}
	-S_{ij}^{(5)} \epsilon_{kli} 
	\int\!\!\int \dVx \dVxp \rho({\bf x} )
	\rho({\bf x'}) 
	\frac{x_{j}x_{k} {x'\!}_l}{|{\bf x} -{\bf x'}|^3} \biggr) \ , 
\\ \nonumber \\ \nonumber \\ 
\int \dVx  \rho v_i \left[ \seven h_{ij}
	\left( \partial_t v_j+v_k \partial_k v_j \right) \right]
	 &=& -{16 \over 45} G\epsilon_{ikl} S_{lj}^{(4)} \int \dVx  \rho x_k 
	{d (v_i v_j) \over dt} 
	 = -{16 \over 45} G \epsilon_{ikl} S_{lj}^{(4)} {d \over dt} 
	\int \dVx  \rho x_k v_i v_j 
\nonumber \\
	 &=& \frac{16}{45} G\biggl( S_{ij}^{(2)}S_{ij}^{(4)} 
	-S_{ij}^{(4)} \epsilon_{kli} \frac{d}{dt} 
	\int\!\!\int \dVx \dVxp 
	\rho({\bf x} )\rho({\bf x'}) 
	\frac{x_{j}x_{k}{x'\!}_l}{|{\bf x} -{\bf x'}|^3} \biggr) \ , 
\\ \nonumber \\ \nonumber \\ 
\label{ie4}
	\int \dVx \rho v_i v_j v_k \partial_k (\seven h_{ij}) &=& 0 \ .
\end{eqnarray}
In deriving (\ref{ie1})--(\ref{ie4}) we have made use of the Newtonian
continuity and Euler equations as well as of the relation
[cf. equation (\ref{s1_1}) in Appendix A]

\begin{equation}
S_{ij}^{(1)}=\int \dVx  \rho 
	\epsilon_{kl(i}v_{j)}x_kv_l 
	+\int\!\!\int \dVx \dVxp \rho({\bf x} )\rho({\bf x'}) 
	\frac{\epsilon_{kl(i} x_{j)}x_k {x'\!}_{l}}{|{\bf x} -{\bf x'}|^3}\ . 
\end{equation}

	Grouping all the terms, we therefore obtain 

\vbox{
\begin{eqnarray}
\label{n_g_de}
\frac{dE}{dt} &=&\frac{16}{45} {G \over c^7}\Bigl( 
	-2 S_{ij}^{(1)}S_{ij}^{(5)}-S_{ij}^{(2)}S_{ij}^{(4)} 
	+2S_{ij}^{(5)} \epsilon_{kli} \int\!\!\int \dVx \dVxp 
	\rho({\bf x} )\rho({\bf x'}) 
	\frac{x_{j}x_{k}{x'\!}_l}{|{\bf x} -{\bf x'}|^3} 
\nonumber\\ 
	&&\hskip 3cm
	+ 2 S_{ij}^{(4)} \epsilon_{kli} \frac{d}{dt} \int\!\!\int \dVx \dVxp 
	\rho({\bf x} )\rho({\bf x'}) 
	\frac{x_{j}x_{k}{x'\!}_l}{|{\bf x} -{\bf x'}|^3} \Bigr) 
\nonumber \\
	&=&-{16 \over 45}{G \over c^7}S_{ij}^{(3)}S_{ij}^{(3)}
	+{16 \over 45}{G \over c^7}
        {d \over dt}\biggl[-2S_{ij}^{(1)}S_{ij}^{(4)}
	+S_{ij}^{(2)}S_{ij}^{(3)}
	+2S_{ij}^{(4)}	\epsilon_{kli} \int\!\!\int \dVx \dVxp 
	\rho({\bf x} )\rho({\bf x'}) 
	\frac{x_{j}x_{k}{x'\!}_l}{|{\bf x} -{\bf x'}|^3} 
	\biggr] \ . 
\end{eqnarray}}

	As done in Section \ref{b_g}, we now assume quasi-periodicity
in the mass-current quadrupole moments and average expression
(\ref{n_g_de}) over several periods. This allows us to discard the
total time derivative term and finally obtain the required result
(\ref{dedteq}).

\subsection{Rate of Angular Momentum Loss} 
\label{am_l}

	Using equation(\ref{n_g_f35}), equation($\ref{dL}$) leads to
the following terms:

\begin{eqnarray}
\int \dVx  \rho \epsilon_{ijk}x_j \partial_k \nine \alpha = 
	-&& \hskip -1.0truecm
	\frac{16}{45} G\int\!\!\int \rho({\bf x} )\rho({\bf x'}) 
	\Bigl( S_{bl}^{(4)} x_b {x'\!}_i \frac{x_l-{x'\!}_l}
	{|{\bf x} -{\bf x'}|^3} 
	+\epsilon_{ijk}\epsilon_{mab} S_{bk}^{(4)} 
	\frac{x_j x_m {x'\!}_a}{|{\bf x} -{\bf x'}|^3} \Bigr) \ , 
\\
	\int \dVx  \rho \epsilon_{ijk} x_j 
	\Bigl[ \partial_t (\seven h_{kl}v_l) 
	+ \partial_m (\seven h_{kl}v_l) v_m \Bigr] &=& 
\nonumber\\
	\hskip -1.0truecm 
	\frac{16}{45} G\biggl[ 2\epsilon_{ijk}S_{bj}S_{bk}^{(5)} 
	+ \int \dVx  \rho && \hskip -1.0truecm
	\Bigl( |{\bf x} |^2 v_l S_{li}^{(5)} 
	- x_k x_b v_i S_{bk}^{(5)} 
	-x_b v_l v_i S_{bl}^{(4)}+x_m v_m v_l S_{il}^{(4)} \Bigr) 
\nonumber\\ 
\label{eq28} 
	+ \int\!\!\int \rho({\bf x} )\rho({\bf x'}) 
	\Bigl(&& \hskip -1.0truecm 
	x_i x_b \frac{x_l-{x'\!}_l} 
	{|{\bf x} -{\bf x'}|^3} S_{bl}^{(4)}
	-|{\bf x} |^2\frac{x_l-{x'\!}_l}{|{\bf x} -{\bf x'}|^3} 
	S_{il}^{(4)} 
	-\epsilon_{ijk}\epsilon_{lab}\frac{x_j x_a {x'\!}_l}
	{|{\bf x} -{\bf x'}|^3} 
	S_{bk}^{(4)} \Bigr) \biggr]\ , 
\\
	\frac{1}{2} \int \dVx  \rho({\bf x} ) 
	\epsilon_{ijk}x_jv_lv_m \partial_k \seven h_{lm} 
	&=&\frac{16}{45} G \int \dVx  \rho (x_b v_i v_m 
	S_{bm}^{(4)}-x_l v_l v_m S_{im}^{(4)}) \ , 
\end{eqnarray}
where we used the Newtonian equation of motion to derive the
right-hand side of equation (\ref{eq28}).

	Grouping all the terms, we then obtain 

\vbox{
\begin{eqnarray}
\label{n_g_dsi}
\frac{dS_i}{dt}&=&\frac{16}{45} {G \over c^7}
        \Bigl[ -2\epsilon_{ijk}	S_{bj}S_{bk}^{(5)} 
	+\int \dVx \rho \Bigl\{
	(x_k x_b v_i S_{bk}^{(5)}-|{\bf x} |^2 v_l S_{li}^{(5)}) 
	+2(x_b v_l v_i S_{bl}^{(4)} - x_m v_m v_l S_{li}^{(4)} ) \Bigr\} 
\nonumber\\
	&& \ \ \ \ \ 
	+ \int\!\!\int \dVx  \dVxp \rho({\bf x} )
	\rho({\bf x'}) \Bigl(-\frac{x_b x_l (x_i-{x'\!}_i)}
	{|{\bf x} -{\bf x'}|^3} S_{bl}^{(4)} + |{\bf x} |^2 
	\frac{(x_l-{x'\!}_l)}{|{\bf x} -
	{\bf x'}|^3} S_{li}^{(4)} \Bigr) \Bigr] 
\nonumber\\
	&=&\frac{16}{45} {G \over c^7}\Bigl[ -2\epsilon_{ijk}
	S_{bj}S_{bk}^{(5)} 
	+\frac{d}{dt}\int \dVx \rho 
	(x_k x_b v_i S_{bk}^{(4)}-|{\bf x} |^2 v_l S_{li}^{(4)}) \Bigr] .  
\end{eqnarray}}

	Once again, we can average equation (\ref{n_g_dsi}) over
several periods and obtain the required result (\ref{djdteq}).

\section{$(0+3.5)$ PN Hydrodynamic Equations}
\label{heqs}

	In this Section we present the final set of $(0+3.5)$ PN
hydrodynamical equations in which the $3.5$ PN radiation-reaction
forces depend {\sl only} on a time-varying mass-current quadrupole
moment. We will present them in a general gauge first and then
distinguish the expressions resulting from Blanchet's gauge and from
our new gauge.

	The general expressions (\ref{hydr1}) and (\ref{hydr3}) for
the continuity and Euler equations can be rewritten as

\begin{eqnarray}
\label{cont_eq}
{\partial \rho \over \partial t}+
	{\partial (\rho v_i) \over \partial x_i}&=&0 \ ,
\\ \nonumber \\
{\partial (\rho w_k) \over \partial t}+
	{\partial (\rho w_k v_i) \over \partial x_i} &=&
\label{euler1}
	-{\partial P \over \partial x_k}-
	\rho{\partial [\phi+\epsilon (\nine\alpha)] \over \partial x_k}
	+\epsilon \rho w_j {\partial (\eight \beta_j) \over \partial x_k} 
	+\epsilon {\rho w_i w_j \over 2}
	{\partial (\seven h_{ij}) \over \partial x_k} \ ,
\end{eqnarray}
where

\begin{equation}
\label{equuu}
w_k \equiv c u_k 
	=\epsilon (\eight \beta_k) +v_j [\delta_{jk} + 
	\epsilon (\seven h_{jk})] \ ,
\end{equation}
and where we define $\epsilon=c^{-7}$ to highlight the
radiation-reaction contributions. Note that the left hand side of
equation (\ref{euler1}) contains a partial time derivatives of $w_k$
rather than of $v_k$. Doing this removes the partial time derivatives
of $\eight \beta_k$ and $\seven h_{jk} v_j$ from the right hand side
[cf. equation (\ref{f35})]. Moreover, since $w_j-v_j=O(\epsilon)$, all
the quantities $w_j$ on the right hand side of equation (\ref{euler1})
can be replaced by the equivalent quantities $v_j$, whenever this is
numerically more convenient.

	In a similar way, the energy equation (\ref{hydr2}) can be
rewritten as

\begin{equation}
{\partial (\rho \varepsilon) \over \partial t}
	+{\partial (\rho \varepsilon v_i) \over \partial x_i}=
	-P{\partial v_i \over \partial x_i} \ ,
\end{equation}
or, if we define ${\bar E} \equiv \varepsilon + \frac{1}{2} w_k w_k$, in the
equivalent form

\begin{eqnarray}
\label{energy}
{\partial (\rho {\bar E}) \over \partial t}
	+{\partial (\rho {\bar E} +P)v_i \over \partial x_i}  = 
	\hskip 13 truecm
\nonumber \\
	 -\rho v_i {\partial [\phi+\epsilon (\nine\alpha)]\over \partial x_i}
	-\epsilon (\eight \beta_j + \seven h_{ij} v_i)
	\biggl({\partial P \over \partial x_j} 
	+\rho {\partial \phi \over \partial x_j} \biggr)
	+\epsilon \rho v_i v_j {\partial (\eight \beta_j) \over \partial x_i}
	+\epsilon {\rho v_i v_j v_k \over 2}
	{\partial (\seven h_{jk}) \over \partial x_i}+O(\epsilon^2) \ .
\nonumber \hskip -1.0truecm \\
\end{eqnarray}

	In the specific case of an equation
of state 
\begin{equation}
\label{eos}
P=(\Gamma-1) \rho \varepsilon \ ,
\end{equation}
the energy equation (\ref{energy}) can be written in a (third) simpler
form

\begin{equation}
{\partial e \over \partial t}+
	{\partial (e v_i) \over \partial x_i}=0 \ ,
\end{equation}
where $e=(\rho \varepsilon)^{1/\Gamma}$. 
 
	As discussed in Sections \ref{b_g} and \ref{n_g}, the metric
coefficients $\nine\alpha$, $\eight \beta_k$ and $\seven h_{ij}$
appearing in (\ref{euler1})--(\ref{energy}) represent the
radiation-reaction potentials and their expressions vary according to
the gauge assumed. In particular, they have the form

\begin{center}
\begin{eqnarray}
\left.
\begin{array}{lccl}
{\mathcal New\ Gauge} 
&\ \ \ \  &\ \ \ \  & 
{\mathcal Blanchet's\ Gauge}  
\\ \\ 
\displaystyle \nine\alpha = \frac{1}{2}(\seven h_{ij}) 
	\left(x_i \partial_j \phi + \partial_j P_i \right)\ , 
& & &
\nine \alpha = 0 \ ,
\\ \\
\eight \beta_k = 0 \ ,
& & &
\displaystyle \eight \beta_k=
	{16 \,G \over 45} \epsilon_{ijk}x_i x_l S_{jl}^{(5)} \ ,
\\ \\
\displaystyle 
\seven h_{ij}=-{32 \,G \over 45} x_k \epsilon_{kl(i} S_{j)l}^{(4)} \ , 
& & & 
\seven h_{ij}=0 \ ,
\end{array} \right.
\nonumber \\
\end{eqnarray}
\end{center}
where $P_i$ is the solution of equation (\ref{pi_eq}). We stress that
the most important difference in the two gauges is given by the
appearance of a fourth or of a fifth time derivative of $S_{ij}$. Note
also that, in the new gauge, both the last term of equation
(\ref{energy}), as well as terms including $\eight \beta_k$, are
zero. 

	Finally, the set of hydrodynamical equations is closed by the
Poisson's equation for the Newtonian gravitational potential $\phi$

\begin{equation}
\label{poisson}
\Delta \phi = 4 \pi G \rho \ .
\end{equation}
In the case of the new gauge, equation (\ref{poisson}) needs to be
supplemented by three additional elliptic equations for the components
of the vector potential $P_i$

\begin{equation}
\label{elliptic_pi}
\Delta P_i = -4 \pi G \rho x_i\ .
\nonumber
\end{equation}
Boundary conditions at $r \rightarrow \infty$ for the linear elliptic
equations (\ref{poisson}) and (\ref{elliptic_pi}) are given by

\begin{equation}
\phi(r) = \frac{G}{r}\int \dVx \rho + O(r^{-3})\ , 
\hskip 2.0truecm
P_i = \frac{G x_k}{r^3}\int \dVx x_k x_i \rho + O(r^{-3})\ .
\end{equation}

	Computing the amplitude and waveforms of the gravitational
waves emitted is clearly of great interest since they provide the
contact with the observations and can be used to extract astrophysical
information about the source. In the wave zone (Thorne 1980) and at a
distance $r=|{\mathbf x}|$ from the source (where $r \gg L$, with
$L$ being the size of the source) the gravitational wave field is
described by the transverse-traceless (TT) part of the linear 3-metric
perturbations (Thorne 1980; Blanchet 1993, 1997).

\begin{equation}
\label{hij_tt}
({\bar h}^{ij}_{(1)})^{\rm TT}(t,\,{\mathbf x}) = 
	- \frac{4G}{c^4}\sum^{\infty}_{l = 2}
	\left\{
	{(-1)^{l} \over l!}\frac{1}{r}\partial_{L-2}
	\left[ M^{(2)}_{ijL-2}
	\left(t- {r \over c} \right) \right] 
  	+ 2 {(-1)^{l} l \over (l +1)!}\frac{1}{r}
	\partial_{aL-2} \left[\epsilon_{ab(i} S^{(1)}_{\!j)bL-2}
	\left(t- {r \over c} \right) \right] 
	\right\}^{\rm TT} +  O(r^{-2})\ ,
\end{equation}
where $\partial_L =\partial_{i_1} \partial_{i_2}\ldots \partial_{i_l}$
and $M_{L}(t-r/c)$, $S_{L}(t-r/c)$ are the ($L=i_1 i_2\ldots i_l$)-th
mass multipole moment and mass-current multipole moment
respectively. The superscript TT refers to projecting out the
transverse, traceless part:

\begin{equation}
[A_{ij}]^{\rm TT} \equiv P_{il} P_{jm} A_{lm} - 
                \frac{1}{2} P_{ij} P_{lm} A_{lm} \ , 
\end{equation}
where $P_{ij} \equiv \delta_{ij} - n_i n_j$ is the projection operator
and $\delta_{ij}$ the usual Kronecker-delta symbol. Restricting our
attention to the contribution given by mass-current quadrupoles,
equation (\ref{hij_tt}) then gives

\begin{equation}
\label{hij_tt_mc}
({\bar h}^{ij}_{(1)})^{\rm TT} = 
	- \frac{8G}{3c^5} \left[ \epsilon_{a b (i} S^{(2)}_{j)a} - 
	\epsilon_{ab(i} n_{j)} n_k S^{(2)}_{ka} 
	\right] \frac{n_b}{r}\ .
\end{equation}

	Note that at the 3.5 PN order, there are no gravitational-wave
``tail effects'' and therefore the mass-current quadrupole $S_{ij}$
corresponds to the gravitational wave moment observed. The usual
states of polarization of the gravitational waves emitted,
$h_+(\theta,\,\varphi)$ and $h_{\times}(\theta,\,\varphi)$ at a
coordinate position $(\theta,\,\varphi)$ on the 2-sphere of radius
$r$, can be derived from (\ref{hij_tt_mc}) after selecting the
orientation of the source and thus the direction of propagation of the
waves (Rasio and Shapiro 1994, Ruffert, Janka and Sch\"afer 1996).

\section{Timescales and Rescaling} 
\label{timescales}

	As mentioned in Section \ref{intro}, we here further explore
the possibility of using the set of $(0+3.5)$ PN hydrodynamical
equations presented above to investigate the onset and growth of the
\hbox{$r$-mode} instability. In particular, we want to address the
problem of the timescales and propose a strategy to suitably rescale
them.

	It is commonly assumed that the evolution of the
\hbox{$r$-mode} instability in a unmagnetized, rotating neutron star
proceeds through three stages (Owen et al. 1998). During the {\sl
initial~} stage, any infinitesimal (axial) perturbation grows
exponentially in a timescale $\tau_{_{\rm GR}}$, set by gravitational
radiation-reaction. This is followed by the {\sl intermediate} stage
during which the amplitude of the mode saturates due to (not yet well
understood) nonlinear hydrodynamic effects; the star is progressively
spun-down as a result of the angular momentum loss via gravitational
waves. The {\sl final} stage of the evolution occurs when the star's
angular velocity is so small that viscous dissipative effects dominate
the radiation-reaction forces and the \hbox{$r$-mode} oscillations are
damped out. The first stage, for a $\ell=m=2$ mode, has been estimated to
be of the order of a few seconds for a neutron star initially rotating
at the break-up limit for several different equations of state, while
the second to be of the order of about one year (Lindblom, Owen and
Morsink 1998).
        
        One complication in simulating \hbox{$r$-mode} oscillations is
that the ``natural'' timescale $\tau_{_{\rm GR}}$ for the instability
to grow and saturate is likely to be much longer than the timescale
over which a numerical computation can be carried out. Even the most
sophisticated three-dimensional Newtonian numerical codes suffer from
numerical viscosity and are able to preserve accurate configurations
only for a limited number of stellar rotations (i.e. $\lesssim 10 -
100$) and this might well be insufficient for the instability to
saturate. Below we review the relevant timescales for the
\hbox{$r$-mode} instability and propose a strategy whereby, with
suitable scaling, we can achieve these timescales in a numerical
simulation. Our brief review follows closely the results presented by
Lindblom, Owen and Morsink (1998).

        Perturbation analysis can be used to estimate ${\tau}_{_{\rm
GR}}$ by assuming that the rate of energy loss to gravitational
radiation emission grows according to

\begin{equation}
\left({d\widetilde{E}\over dt}\right)_{_{\rm GR}} = 
        -\frac{2\widetilde{E}}{{\tau}_{_{\rm GR}}} \ ,
\end{equation}
where $\widetilde{E}>0$ is the energy in the mode and ${\tau}_{_{\rm
GR}}<0$. In the corotating frame of the equilibrium unmagnetized star,
$\widetilde E$ can be calculated as

\begin{equation}
\label{e_inmode}
\widetilde{E}={1\over 2} 
        \int\left[ \rho\, \delta {\mathbf v}\cdot \delta{\mathbf v}\,^*
        + \left({\delta p\over\rho} - \delta\phi\right)  
        \delta \rho^*\right]\dVx \ .
\end{equation}

        The lowest order expressions for the Eulerian density
perturbation $\delta \rho$ and velocity perturbations $\delta v^a$ can
be deduced from the perturbed fluid equations and, in a spherical
coordinate system $(r,\;\theta,\;\varphi)$, have the form (Lindblom,
Owen and Morsink 1998; Lindblom, Mendell and Owen 1999)

\begin{eqnarray}
\label{drho}
{\delta \rho\over\rho} =
	\frac{(2\ell+1)}{\ell(\ell+1)\sqrt{2\ell+3}} 
	\left(\frac{\alpha_{_A} R^2\Omega^2}{\sqrt{2\ell+3}} \right)
	{d\rho\over dp} 
        \left[{2 \ell\over 2\ell+1}\sqrt{\ell\over \ell+1}
        \left({r\over R}\right)^{\ell+1}+\delta\Psi(r)\right] Y_{\ell+1\,\ell}
        \,e^{i\omega t} \ ,
\end{eqnarray}
where $\delta\Psi(r)$ is proportional to the gravitational potential,
and the axial velocity perturbations are given by

\begin{equation}
\label{dv}
\delta {\mathbf v} = \alpha_{_A} R \Omega \left({r\over R}\right)^\ell
	{\mathbf Y}^{B}_{\ell\,m} e^{i\omega t} \ .
\end{equation}
Here, $R$ and $\Omega$ are the radius and angular velocity of the
unperturbed star, $\alpha_{_A}(t)$ is a dimensionless coefficient
parameterizing the amplitude of the perturbation, $\omega$ is the
(Eulerian) frequency of the mode, and ${\mathbf Y}^{B}_{\ell\,m}$ is the
magnetic-type vector spherical harmonic. Given an axial perturbation
with periodic dependence $e^{i (m \varphi + \omega t )}$ and the
definition (\ref{e_inmode}) of the energy in the mode, the perturbed
fluid equations yield the following general expression for the time
derivative of $\widetilde{E}$ (Ipser and Lindblom 1991; Lindblom, Owen
and Morsink 1998)

\begin{eqnarray}
\label{de_general}
{d\widetilde{E}\over dt} =
        -\int\left(2\eta\delta\sigma^{ab}\delta\sigma_{ab}^*
        +\zeta \delta\sigma \delta\sigma^*\right)\dVx
         -\omega(\omega+m\Omega)\sum_{\ell\geq 2} N_\ell \omega^{2\ell}
        \left(|\delta I_{\ell m}|^2+\frac{|\delta S_{\ell m}|^2}{c^2}\right) 
        \ ,
\end{eqnarray}
where $\eta$ and $\zeta$ are the shear and bulk viscosities of the
fluid (taken as given functions of the density and temperature) and

\begin{equation}
N_\ell = {4\pi G\over c^{2\ell+1}} {(\ell+1)(\ell+2)\over \ell(\ell-1)[(2\ell+1)!!]^2} \ .
\end{equation}

      It is easy to distinguish in expression (\ref{de_general}) the
suppressing viscous terms, driven by the perturbed shear $\delta
\sigma_{ab}$ and expansion $\delta\sigma$, from the driving
gravitational radiation terms, driven by the time-varying mass $\delta
I_{\ell m}$ and mass-current $\delta S_{\ell m}$ multipole moments of the
perturbed fluid\footnote{Note that $\omega(\omega+m\Omega) < 0$.}. The
explicit contribution to the imaginary part of the frequency of the
mode due to gravitational radiation-reaction can then be calculated as
(Lindblom, Owen and Morsink 1998)

\begin{equation}
\label{tau_gr_1}
{1\over \tau_{_{\rm GR}}} =- {1\over 2 \widetilde{E}} 
        \left( {d\widetilde{E}\over dt} \right)_{_{\rm GR}}
        =-{32\pi G\Omega^{2\ell+2} \over c^{2\ell+3}}
        {(\ell-1)^{2\ell}\over [(2\ell+1)!!]^2}
        \left({\ell+2\over \ell+1}\right)^{2\ell+2}
        \int_0^R\rho\,r^{2\ell+2} dr \ .
\end{equation}
where, as first pointed out by Papaloizou and Pringle (1978), we have
used the following relation between the frequency of the mode and the
star angular velocity

\begin{equation}
\label{omegas}
\omega = \omega_{\rm rot} - m \Omega =
        \frac{2 m \Omega}{\ell(\ell+1)} - m \Omega =
         - {(\ell-1)(\ell+2)\over \ell+1}\Omega \ .
\end{equation}
Here $\omega_{\rm rot}$ is the angular frequency of the mode in the
corotating frame and the last expression in (\ref{omegas}) refers to
the case in which $m=\ell$. Note that the contribution to the growth rate
in (\ref{tau_gr_1}) comes solely from the current multipole moments
$\delta S_{\ell\,\ell}$ since we have implicitly neglected the contributions
coming from the mass multipole moments $\delta I_{\ell\,\ell}$. Such an
approximation is reasonable because the density perturbations are one
order in $\Omega$ higher than the correspondent velocity perturbations
and because the density perturbations generate gravitational radiation
at a higher frequency (Lindblom, Owen and Morsink 1998).

        The general expression for $\tau_{_{\rm GR}}$ in
(\ref{tau_gr_1}) can be rewritten in a number of alternative ways, some
of which are more useful within a computational context. Depending on
whether the sequence of initial data is specified in terms of the
ratio $R/M$, or in terms of the mass $M$, or of the angular velocity
$\Omega$, we can rewrite (\ref{tau_gr_1}) respectively as

\begin{mathletters}
\begin{eqnarray}
\label{tau_gr_2}
\tau_{_{\rm GR}} 
        &=& - A_\ell \frac{c^{2\ell+3}}{G} \frac{R^{2\ell+3}}{\cal I}
        \frac{1}{(\Omega M)^{2\ell+2}} \left(\frac{M}{R}\right)^{2\ell} M \ ,
        \\ \nonumber \\ 
\label{tau_gr_3}
        &=& - A_\ell \frac{c^{2\ell+3}}{G^{\ell+2}} \frac{R^{2\ell+3}}{\cal I}
        \left(\frac{\Omega_{_K}}{\Omega}\right)^{2\ell+2} 
        \left(\frac{R}{M}\right)^{\ell+3} M \ ,
        \\ \nonumber \\ 
\label{tau_gr_4}
        &=& - A_\ell \left(\frac{c^2}{G}\right)^{\ell+3/2} 
	\frac{R^{2\ell+3}}{\cal I}
        \left(\frac{\Omega_{_K}}{\Omega}\right)^{2\ell+1} 
        \left(\frac{R}{M}\right)^{\ell+3/2} \frac{1}{\Omega} \ ,
\end{eqnarray}
\end{mathletters}
where

\begin{equation}
\label{A_l}
A_\ell \equiv \frac{1}{24}
        \frac{[(2\ell+1)!!]^2}{(\ell-1)^{2\ell}}
        \left(\frac{\ell+1}{\ell+2}\right)^{2\ell+2} \ , 
        \hskip 2.0 truecm 
\Omega_{_K} = \sqrt \frac{GM}{R^3}      
\end{equation}
and 

\begin{equation}
\label{calI}
{\cal I} \equiv \frac{1}{\bar \rho} \int_0^R\rho\,r^{2\ell+2} dr \ ,
        \hskip 2.0 truecm 
{\bar \rho} \equiv \frac{3M}{4 \pi R^3} \ .
\end{equation}

        In general, the integral in (\ref{calI}) needs to be computed
numerically, but, in the case of a polytropic equation of state
$P=K\rho^{\Gamma}$ with $\Gamma = 2$, it can be computed analytically
and, in particular, (Jeffrey 1995)

\begin{equation}
\label{def_int}
\frac{\cal I}{R^{2\ell+3}} 
         =  \frac{1}{3\pi^{2\ell+1}}
        \int_0^{\pi} \sin x \,x^{2\ell+1} dx 
         =  \frac{(2\ell+1)!}{3\pi^{2\ell+1}}
        \left\{\sum_{k=0}^\ell (-1)^{k+2} 
        \frac{\pi^{2\ell-2k+1}}{(2\ell-2k+1)!} \right\} \ .
\end{equation}

        Suppose we now impose the condition that the {\it
computational timescale} $\tau_{_C}$, expressed as a multiple $N$ of
the stellar rotations, be identical to the growth-time $\tau_{_{\rm
GR}}$

\begin{equation}
\label{tau_c}
\tau_{_C} \equiv N \frac{2 \pi}{\Omega} = |\tau_{_{\rm GR}}|
\end{equation}
Using (\ref{def_int}) and (\ref{tau_gr_4}), expression
(\ref{tau_c}) then becomes a condition on the ratio $c^2 R/(GM)$, i.e.

\begin{equation}
\label{condition}
\frac{c^2 R}{G M} = \left[
        \frac{2 \pi N}{A_\ell} \frac{\cal I}{R^{2\ell+3}}
        \left(\frac{\Omega}{\Omega_{_K}}\right)^{2\ell+1} 
        \right]^{2/(2\ell+3)}  
        \simeq 0.68 \ N^{2/7} \ ,
\end{equation}
where the numerical coefficient comes from considering $\ell=2$, $\Gamma
= 2$ and $\Omega_{_K} = \Omega$. According to (\ref{condition}), it is
always possible to rescale the value of the constant $c$ in such a way
as to make the growth-time compatible with the timescale over which
the numerical computations can be carried out. Provided we maintain
the inequality $\tau_{_C} = | \tau_{_{\rm GR}} | \gg \Omega^{-1}$ or
$N \gg 1$, this rescaling should in no way affect the profiles of the
physical parameters during the evolution, but only shorten the
evolution time over which their growth and saturation occur.
Alternatively, one can choose $M/R$ to be sufficiently large to reduce
the growth-time in accord with (\ref{tau_gr_3}) and then scale the
results to stars with more realistic compaction ratios. A similar
rescaling technique has also been adopted to accelerate the cooling of
a hot neutron star and study its collapse to a black hole (Baumgarte,
Shapiro and Teukolsky 1996).

\section{Conclusions}
\label{conclusion}

        We have presented a new set of $(0+3.5)$ PN hydrodynamical
equations in which a $3.5$ PN radiation-reaction force due to a
time-varying mass-current quadrupole moment is considered. Within this
system of equations, the hydrodynamics is essentially Newtonian except
for the inclusion of the relativistic nonconservative effects related
to the emission of gravitational radiation.

        We have cast this set of equations in a form which is suitable
for numerical implementation. In the alternative 3.5 PN approach by
Blanchet (1993, 1997), the radiation-reaction terms depend on the
fifth time derivative of the mass-current quadrupole moment
$S_{ij}$. Evaluating such a term accurately within a standard second
order numerical scheme could pose a problem. Instead, we have chosen a
particular gauge in which the radiation-reaction effects depend at
most on the fourth time derivative of $S_{ij}$ and can therefore be
calculated accurately. The additional complication that arises with
this gauge choice are three linear, elliptic equations for the
components of a vector potential. The solution of such equations is no
more difficult than the solution of Poisson's equation for the
Newtonian gravitational potential and can be performed in an identical
fashion.

	Simulating the onset and growth of the \hbox{$r$-mode}
instability in rotating neutron stars is highly desirable. A
$(0+3.5)$PN approach may be considerably simpler than a fully general
relativistic one and allows one to neglect all conservative
relativistic effects, which should be perturbative, and focus
exclusively on the radiation-reaction effects due to a time-varying
mass-current quadrupole moment. We have also proposed a suitable
rescaling that will make the timescale for the onset and saturation of
the \hbox{$r$-mode} instability compatible with any reasonable
integration time imposed by computational constraints. Work is
presently in progress to implement these equations in a numerical code
(Ruffert et al. 1999).

\acknowledgments 

We are grateful to L. Blanchet for his helpful comments and for
carefully reading the manuscript. H. Asada would like to thank
Y. Kojima for useful conversations. This work was supported by NSF
Grant AST 96-18524 and NASA Grant NAG 5-7152 at Illinois and a JSPS
Fellowship to M.~Shibata for Research Abroad. M. Shibata also
acknowledges the kind hospitality of the Department of Physics of the
University of Illinois at Urbana-Champaign.

\section*{\label{A} \break \break Appendix A \break \break
	Strategy for computation of $S_{ij}^{(n)}$}

	In this Section we present the analytic integral expressions
for the first, the second, and the third time derivative of the mass
current quadrupole moment $S_{ij}$. The expressions derived here can
then be used, after taking numerical time derivatives, to compute
$S_{ij}^{(4)}$ and (if necessary) $S_{ij}^{(5)}$.

	The first time derivative of $S_{ij}$ is easily derived from
(\ref{Sij}), after setting $v_i = dx_i/dt$, to yield

\begin{mathletters}
\begin{eqnarray}
\label{s1_1}
S_{ij}^{(1)}&&=\epsilon_{kl(i} \int \dVx  \rho
	\biggl(v_{j)} v_l + x_{j)} {dv_l \over dt}\biggr)x_k + O(\epsilon)\ ;
\\
\label{s1_2}
	&&=\epsilon_{kl(i}\int \dVx  \rho 
	\left( v_{j)}v_l - x_{j)} \partial_l \phi \right)x_k + O(\epsilon)\ .
\end{eqnarray}
\end{mathletters}
In deriving (\ref{s1_2}) we have used the continuity and the Newtonian
Euler equation,

\begin{equation}
{\partial v_i\over \partial t}+ v_j {\partial v_i \over \partial x_j}
	= -\frac{1}{\rho}{\partial P \over \partial x_i}- {\partial
	\phi \over \partial x_i} \ ,
\end{equation}
and exploited the following identity

\begin{equation}
\epsilon_{km(i} \int \dVx   x_{j)} x_k \partial_m P =0 \ . 
\end{equation}

	A similar procedure is used for the second time derivative,
which can be written as

\begin{mathletters}
\label{s2}
\vbox{
\begin{eqnarray}
\label{s2_1}
S_{ij}^{(2)} &=& \epsilon_{kl(i} \Bigg\{ \int \dVx P
	x_k \left[ \partial_{j)} v_l + \partial_l v_{j)} \right] - 
	\int \dVx \rho x_k \left[ (\partial_{j)} \phi) v_l 
	+ 2 v_{j)} \partial_l \phi \right] 
\nonumber \\
	&& \qquad \ \ - \int \dVx \rho \left[ x_{j)} v_k \partial_l \phi
	+ x_{j)} x_k \left[\partial_l (\partial_t \phi) + 
	v_m \partial_{lm}\phi \right] \right] \Bigg\} + O(\epsilon) \ ;
\\
\label{s2_2}
	& = &  \epsilon_{kl(i} \Bigg\{ 
	\int d^3 {\bf x}\, P x_k 
	\left[ \partial_{j)} v_l + \partial_l v_{j)} \right]  + 
	\int d^3 {\bf x}\, \nabla \cdot (\rho {\mathbf v}) 
	\left[ x_{j)} x_k \partial_l \phi \right]
\nonumber \\
	& & \qquad \ \ - \int d^3 {\bf x}\, \rho x_k 
	\left[ x_{j)} \partial_{l} (\partial_t \phi)  + 
	v_{l} \partial_{j)} \phi +
	v_{j)}\partial_l \phi \right] \Bigg\} + O(\epsilon) \ . 
\end{eqnarray}}
\end{mathletters}

Note that we have proposed two different expressions for
$S_{ij}^{(2)}$, where the second one [i.e. (\ref{s2_2})] made use of
the following identity

\begin{eqnarray}
\int \dVx \rho \left[ v_{(i} \partial_{j)} \phi + v_k x_{(i} 
	\partial_{j)k}\phi\right] = 
	\int \dVx \rho v_k \partial_k 
	\left[ x_{(i} \partial_{j)} \phi \right] = && \hskip -0.6truecm 
	\int \dVx \partial_k 
	\left[\rho v_k x_{(i} \partial_{j)} \phi \right] 
	-\int \dVx \left[x_{(i} \partial_{j)} \phi \right]
	\nabla \cdot (\rho {\mathbf v})
\nonumber \\ 
	= && \hskip -0.6truecm 
	-\int \dVx \left[x_{(i} \partial_{j)} \phi \right]
	\nabla \cdot (\rho {\mathbf v}) \ ,
\end{eqnarray}
in order to eliminate the mixed second partial derivatives of the
gravitational potential.

	Expressions (\ref{s1_1})--(\ref{s2_2}) apply for a generic
equation of state. However, in deriving $S_{ij}^{(3)}$ we will need a
time derivative of the volume integral of the pressure and a specific
equation of state must be specified. In the case of an equation of
state $P=(\Gamma-1) \rho \varepsilon$, we obtain

\begin{mathletters}
\label{s3}
\vbox{
\begin{eqnarray}
\label{s3_1}
S_{ij}^{(3)} &=& \epsilon_{kl(i} \Bigg\{ \int \dVx (\Gamma - 1)  
	\rho \varepsilon \Bigl\{(1 -\Gamma) x_k 
	\left( \partial_{j)} v_l + \partial_l v_{j)} \right) (\partial_m v_m) 
	+v_k \left(\partial_{j)} v_l + \partial_l v_{j)} \right) 
\nonumber \\
	&& \hskip 4.0truecm 
	+x_k \left[\partial_{j)} a_l + \partial_l a_{j)} - 
	(\partial_{j)} v_n)(\partial_n v_l) 
	-(\partial_l v_n) (\partial_n v_{j)}) \right] \Bigr\} 
\nonumber \\
	&& \qquad - \int \dVx \rho \left\{
	x_{j)}a_k \partial_l \phi + 3 v_{j)} v_k\partial_l \phi 
	+ x_k \left[ a_l \partial_{j)}\phi + 2 a_{j)} \partial_l \phi   
	+ v_l \partial_{j)} (\partial_t \phi)  
	+ v_l v_n \partial_{j)n}\phi \right] \right\}
\nonumber \\
	&& \qquad - \int \dVx \rho \left\{
	[3v_{j)} x_k +2x_{j)} v_k] 
	[\partial_l (\partial_t \phi) + v_n \partial_{ln}\phi]
	+ x_{j)}x_k \left[\partial_l \partial_{tt}\phi + 
	2v_n \partial_{ln}(\partial_t \phi) 
	+a_n \partial_{ln}\phi \right] \right\}
\nonumber \\
	&& \qquad + \int \dVx
	x_{j)}x_k \partial_l(\rho v_m v_n)\partial_{mn}\phi \biggr] 
	\Bigg\} +O(\epsilon)\ ,
	\nonumber \\
\end{eqnarray}}
or equally

\vbox{
\begin{eqnarray}
\label{s3_2}
S_{ij}^{(3)} & = &  \epsilon_{kl(i} \Bigg\{ 
        \int d^3 {\bf x}\, P 
        \left[ \partial_{j)} v_l + \partial_l v_{j)} \right]
        \left( v_k - (\Gamma-1)x_k \partial_n v_n \right)
\nonumber \\ 
        & & \qquad \ \ 
	+ \int d^3 {\bf x}\, P x_k 
        \left[\partial_{j)} a_l + \partial_l a_{j)} - 
        (\partial_{j)} v_n)(\partial_n v_l) 
        -(\partial_l v_n) (\partial_n v_{j)}) \right]      
\nonumber \\ 
	& & \qquad \ \ 
	+ \int  d^3 {\bf x}\, \nabla \cdot (\rho {\mathbf v}) 
	\bigg[ \partial_l \phi \left( v_{j)} x_k + x_{j)} v_k \right) 
	+ x_{j)} x_k \left[\partial_{l} (\partial_t \phi) + 
	v_n \partial_{nl}\phi \right]\bigg]
\nonumber \\ 
	& & \qquad \ \ 
	+ \int  d^3 {\bf x}\, x_{j)} x_k \partial_l \phi 
	\left[\partial_n\left(\rho  a_n\right) - 
	\partial_n (\rho v_m \partial_m v_n ) \right] -
	\int  d^3 {\bf x}\, \rho v_k 
	\left[ x_{j)} \partial_{l} (\partial_t \phi) + 
		\left(\partial_{j)} \phi \right) v_l + 
		v_{j)} \partial_{l} \phi \right]
\nonumber \\ 
	& & \qquad \ \ 
	- \int d^3 {\bf x}\, \rho x_k 
	\bigg[ v_{j)} \left[ 2\partial_l (\partial_t \phi) + v_n
		\partial_{nl} \phi \right] + x_{j)}
		\left[ \partial_l (\partial_{tt} \phi) + 
		v_n \partial_{nl} (\partial_t \phi) \right]
\nonumber \\ 
	& & \qquad \qquad \qquad \qquad \ \ \ 
	v_l 
	+ \left[ \partial_{j)} (\partial_t \phi) + v_n\partial_{n j)} \phi
	\right] + a_l \partial_{j)} \phi +  a_{j)} \partial_{l} \phi 
	\bigg] \Bigg\} \ .
\nonumber \\ 
\end{eqnarray}}
\end{mathletters}

\noindent where we have defined

\begin{equation}
a_i \equiv \frac{dv_i}{dt}= 
	-{1 \over \rho}\partial_i P  - \partial_i \phi \ ,
\end{equation}
and used the relation

\begin{equation}
\frac{d}{dt} \int \dVx P = \int \dVx \rho \frac{d}{dt} 
	\left(\frac{P}{\rho}\right) = - \int \dVx P (\Gamma - 1) 
	\partial_j v_j \ .
\end{equation}

	The partial time derivatives of the gravitational potential
$(\partial_t \phi)$ and $(\partial_{tt} \phi)$ appearing in (\ref{s2})
and (\ref{s3}) satisfy the following elliptic equations (Nakamura and
Oohara 1989)

\begin{eqnarray}
&&\Delta (\partial_t \phi) = -4\pi G \partial_k (\rho v_k) \ ,
\\ \nonumber \\
&&\Delta (\partial_{tt} \phi)=4\pi G
[\partial_{ij}(\rho v_i v_j)+\Delta P + 
	\partial_i (\rho \partial_i \phi)] \ , 
\end{eqnarray}
where $\Delta$ denotes the flat spatial Laplacian. While
$S_{ij}^{(4)}$ and $S_{ij}^{(5)}$ could also be expressed through
similar integral expressions, this is not useful in general. For a
numerical method which is accurate in time and space at the order $n$,
the maximum time derivative of $S_{ij}$ which can be calculated
accurately is $(n+1)$. This is because, for a generic
$S_{ij}^{(n+1)}$, we need $n$ spatial derivatives and $n$ partial time
derivatives of $\phi$. As a result, if one is using a numerical method
which is second order accurate in space and time, analytic integral
expressions are reliable at most up to $S_{ij}^{(3)}$. The fourth and
fifth time derivatives need to be obtained by finite differencing of
$S_{ij}^{(3)}$ with increasingly larger truncation errors. This
consideration is the guideline for our formalism, for which we need
only compute $S_{ij}^{(4)}$.

\section*{\label{B} \break \break Appendix B \break \break
	Canonical Form of the Linear Metric}

	Here we obtain the parts of the metric associated with
radiation-reaction potential. We follow the notation of Blanchet
(1993). We first recall that the components of the linearized metric
${\tilde h}^{\mu \nu}_{(1)}$ in canonical form and in the harmonic
gauge condition (Thorne 1980) are given by

\begin{eqnarray}
&&\tilde \alpha=0\ ,
\\
&&\tilde \beta^i={4G \over c^3}\sum^{\infty}_{l=1}{(-1)^l l \over (l+1)!}
	\epsilon_{iab}\partial_{aL-1}
	\biggl[{1 \over r}S_{bL-1}\Bigl(t-{r \over c}\Bigr)\biggr]\ ,
\\
&&\tilde h_{ij}={8G \over c^4}\sum^{\infty}_{l=2}{(-1)^l l \over (l+1)!}
	\partial_{aL-2}\biggl[{1 \over r}\epsilon_{ab(i} S_{j)bL-2}^{(1)}
\Bigl(t-{r \over c}\Bigr)\biggr] \ ,
\end{eqnarray}
where we have considered {\it only} the terms related to mass-current
multipole moments $S_L$. Here $\partial_L =\partial_{i_1}
\partial_{i_2}\ldots \partial_{i_l}$ and $L=i_1 i_2\ldots i_l$ is a
compact expression for $l$ indices. Since there is no ``tail'' term at
the linear order, we can derive the radiation-reaction metric by
taking the half-difference of the retarded and advanced waves
(Blanchet 1993)

\begin{eqnarray}
\label{betai}
&&\tilde \beta^i={4G \over c^3}\sum^{\infty}_{l=1}{(-1)^l l \over (l+1)!}
	\epsilon_{iab}\partial_{aL-1}
	\biggl[{S_{bL-1}(t-r/c)-S_{bL-1}(t+r/c) \over 2r}\biggr]\ ,
\\
&&\tilde h_{ij}={8G \over c^4}\sum^{\infty}_{l=2}{(-1)^l l \over (l+1)!}
	\partial_{aL-2}	\biggl[\epsilon_{ab(i}
	{S_{j)bL-2}^{(1)}(t-r/c)-
	S_{j)bL-2}^{(1)}(t+r/c) \over 2r}\biggr]\ . 
\end{eqnarray}
Expanding the numerators of the right-hand sides with respect to
$c^{-1}$ in order to determine the near zone metric, we obtain
equations (\ref{harmo8}) and (\ref{harmo7}) as the lowest order of the
$l=2$ mode. The gauge transformation necessary in order to set the
``new'' linear shift $\beta^i=0$ is therefore simply given by

\begin{equation}
\label{xi}
\xi^i = - {4G \over c^2}\sum^{\infty}_{l=2}{(-1)^l l \over (l+1)!}
	\epsilon_{iab}\partial_{aL-1}
	\biggl[{S_{bL-1}^{(-1)}(t-r/c)-S_{bL-1}^{(-1)}(t+r/c) 
	\over 2r}\biggr] \ ,
\end{equation}
Using (\ref{xi}), the expression of $\tilde h_{ij}$ in the new gauge
is 

\begin{eqnarray}
\label{hij_n}
h_{ij} &=&\tilde h_{ij} + 2 \partial_{(i} \xi_{j)} = 
	{8G \over c^4} \sum^{\infty}_{l=2} {(-1)^l l \over (l+1)!}
	\frac{l+2}{2l+1}\partial_{aL-2}	\biggl[\epsilon_{ab(i}
	{S_{j)bL-2}^{(1)}(t-r/c)-
	S_{j)bL-2}^{(1)}(t+r/c) \over 2r}\biggr]
\nonumber \\
& & \qquad \qquad \qquad \ \ 
 	- {8G \over c^2}\sum^{\infty}_{l=2}{(-1)^l l \over (l+1)!}
	\epsilon_{ab(i}{\hat \partial}_{j)aL-1}
	\biggl[{S_{bL-1}^{(-1)}(t-r/c)-S_{bL-1}^{(-1)}(t+r/c) 
	\over 2r}\biggr] \ ,
\end{eqnarray}
where ${\hat \partial}_L$ is the (symmetric) trace-free part
of $\partial_L$. Equation (\ref{hij_n}) should be compared with the
equivalent one obtained in Blanchet's gauge (we recall that we
report here only the contributions due to time-varying mass-current
multipoles) [cf. equation (2.8c) of Blanchet 1997] 

\begin{eqnarray}
\label{hij_bg}
\big( h_{ij} \big)^{\rm Blanchet's}_{\rm \ \ gauge} = 
	- {8G \over c^2}\sum^{\infty}_{l=2}{(-1)^l l \over (l+1)!}
	\frac{2l+1}{l-1} \epsilon_{ab(i}{\hat \partial}_{j)aL-1}
	\biggl[{S_{bL-1}^{(-1)}(t-r/c)-S_{bL-1}^{(-1)}(t+r/c) 
	\over 2r}\biggr] \ . 
\end{eqnarray}
Expression (\ref{hij_bg}), as well as the the second term in
(\ref{hij_n}), provide no contribution at 3.5 PN order.

	Using (\ref{betai}) and (\ref{hij_n}) at the lowest order of
the PN expansion of the $l=2$ mode, we obtain equations
(\ref{our_mc}).

\section*{\label{C} \break \break Appendix C \break \break
	Time-Slice condition and equation for \large{$\ \nine\alpha$}}

	We here discuss the choice of a time-slice condition and the
derivation of the elliptic equation (\ref{eqalp}) for $\nine\alpha$.
We start by considering
the evolution equation for the 3-metric $\gamma_{ij}$

\begin{equation}
\label{dtgammaij}
\frac{1}{c}\partial_t \gamma_{ij} = 
	-2 \alpha K_{ij} + D_i \beta_j + D_j \beta_i \ , 
\end{equation}
and the momentum constraint equations

\begin{equation}
\label{mom_c}
D_j K^j_{\ i} - D_i K = \frac{8 \pi G}{c^4}J_i \ ,
\end{equation}
where $D_i$ is the covariant derivative with respect to $\gamma_{ij}$,
$K_{ij}$ is the extrinsic curvature tensor, and $K=K_i^{~i}$. The 
current source term on the left hand side of (\ref{mom_c}) is defined
as $J_{\mu} \equiv - \gamma_{\mu \nu} T^{\nu \alpha} n_{\alpha}$, with
$n^{\mu}=(1/\alpha, -\beta^k/\alpha)$ being the normal to the spatial
slice. The 3.5 PN expressions of (\ref{dtgammaij}) and (\ref{mom_c})
are given respectively by

\begin{equation}
\label{dtgammaij_pn}
\frac{1}{c}\partial_t (\seven h_{ij}) = -2 \eight K_{ij} + 
	\partial_i (\eight\beta_j) + \partial_j (\eight\beta_i) \ , 
\end{equation}
and 

\begin{equation}
\label{mom_c_pn}
\partial_i (\eight K_{ij}) - \partial_j (\eight K) = 0 \ ,
\end{equation}
where $\four J_i = 0$ (Asada, Shibata, and Futamase 1996).  

	Taking a further spatial derivative of (\ref{dtgammaij_pn})
and using the constraints (\ref{mom_c_pn}), we then obtain

\begin{equation}
\label{8keq0}
2 \partial_i (\eight K) = 2 \partial_i (\eight K_{ij}) =  
	\partial_{ii} (\eight\beta_j) + \partial_{ij} (\eight\beta_i) -
	\frac{1}{c}\partial_{ti} (\seven h_{ij})\ . 
\end{equation}

	Performing now an infinitesimal gauge transformation yielding
[cf. equations (\ref{n_g_metric})--(\ref{hij7})]

\begin{eqnarray}
\label{beta8eq0_a}
\eight\beta_i&=&0 \ , 
\\
\label{hij7_a}
\seven h_{ij}&=&-\frac{32 \,G}{45} x_k \epsilon_{kl(i} S_{j)l}^{(4)} \ . 
\end{eqnarray}
will set to zero the right hand side of equation (\ref{8keq0}) [we
recall that $\partial_i (\seven h_{ij})=0$] and thus require $\eight
K$ to be a constant. As a result, we choose as slice condition

\begin{equation}
\label{k0}
K=0\ ,
\end{equation}
at all times. Condition (\ref{k0}) is known as the maximal time-slice
condition (Arnowitt, Deser and Misner 1962; Smarr and York 1978;
Sch\"afer 1983; Blanchet, Damour and Sch\"afer 1990).  As a
consequence of (\ref{k0}), the evolution equation of $K$ is given by

\begin{equation}
\label{dkdt}
\frac{1}{c}\partial_t K = 
	- D_k D^k \alpha + {4\pi G \over c^2} 
	\alpha (\rho_{_E} + S) + \alpha K_{ij}K^{ij} = 0 \ , 
\end{equation}
where $\rho_{_E} \equiv T_{\mu \nu} n^{\mu} n^{\nu}$ and $S \equiv
T_{\mu \nu} \gamma^{\mu \nu}$.

	The metric coefficient $\nine\alpha$ is then obtained as the
solution of the 3.5 PN expression of the elliptic equation
(\ref{dkdt}):

\begin{equation}
\label{alphaeq}
D_k D^k \alpha = {4\pi G\over c^2} 
	\alpha (\rho_{_E} + S)  + \alpha K_{ij}K^{ij}\ .
\end{equation}
After discarding all the 3.5 PN terms except those arising from the
mass-current quadrupole, the 3.5 PN expression of the left-hand side
of equation (\ref{alphaeq}) is written as [cf. equation (\ref{pnexp})]

\begin{equation}
D_k D^k \alpha = \Delta (\nine\alpha) -
	\partial_i (\seven h_{ij}\partial_j \phi ) \ ,
\end{equation}
while the (full) right-hand side of equation (\ref{alphaeq}) is
rewritten as

\begin{equation}
\label{source}
{4\pi G\over c^2} \alpha (\rho_{_E} + S) + \alpha K_{ij}K^{ij} = 
	\frac{4\pi G\alpha}{c^2} 
	\left [\rho \left\{ 1 + \frac{1}{c^2} 
	\left(\varepsilon + \frac{P}{\rho} \right) \right\}
	\left(1 + \frac{2}{c^2} \gamma^{ij}w_iw_j \right) + 
	{2 \over c^2}P \right] + \alpha K_{ij} K^{ij}\ . 
\end{equation}

	It is easy to estimate that, at the 3.5 PN level, the
contributions to expression (\ref{source}) coming from the mass
current quadrupole moment are at most $O(c^{-11})$ and that the
slicing condition for $\nine\alpha$ is therefore given by
(\ref{eqalp}). This follows from the fact that the contribution in
$\rho$, $\varepsilon$, and $P$ is $O(1)$, and is $O(c^{-7})$ for
$\gamma^{ij}w_i w_j$. As a consequence, the contribution of the mass
current quadrupole moment from the terms in the curly brackets of
(\ref{source}) is at most of $O(c^{-9})$. Similar considerations apply
also for the last term in the right-hand side of (\ref{source}) where
$K_{ij}=O(c^{-3})$, while the contribution of the mass-current
quadrupole moment in $K_{ij}$ is $O(c^{-8})$ [cf. equation
(\ref{dtgammaij})]. As a result, the mass-current contribution in
$\alpha K_{ij} K^{ij}$ is at most $O(c^{-11})$.

\end{document}